\documentclass[12pt,a4paper,twocolumn,fleqn]{narms}
\usepackage[latin1]{inputenc}

\usepackage{subfigure}
\usepackage{graphicx}
\DeclareGraphicsExtensions{.png,.pdf}
\usepackage{timesmt}

\usepackage{hyperref}
\hypersetup{
    pdfborder=0 0 0,
    colorlinks=false,
    breaklinks=true,
    urlcolor=false,
    linkcolor=false,
    bookmarksopen=true,
    pdftitle={Metamodel-based importance sampling for the simulation of rare events (FullTextPaper)},
    pdfauthor={V. Dubourg, F. Deheeger, B. Sudret},
    pdfsubject={Metamodel-based importance sampling for the simulation of rare events},
    pdfkeywords={Metamodel-based importance sampling for the simulation of rare events}
  }
\usepackage{color}
\usepackage{colortbl}
\usepackage{multirow}
\makeatletter
\def\hlinewd#1{
\noalign{\ifnum0=`}\fi\hrule \@height #1
\futurelet\reserved@a\@xhline}
\makeatother

\newcommand{\hlineB}{\hlinewd{0.85pt}}
\usepackage{tikz}
\usetikzlibrary{shapes}
\usetikzlibrary{patterns}

\usepackage{amsmath,amssymb,amsfonts,amsbsy,bbm}
\renewcommand{\eqref}[1]{Eq.~(\ref{#1})}

\newcommand{\cc}{{\mathcal C}}
\newcommand{\cd}{{\mathcal D}}

\newcommand{\cl}{{\mathcal L}}

\newcommand{\cn}{{\mathcal N}}

\newcommand{\cp}{{\mathcal P}}

\newcommand{\cx}{{\mathcal X}}

\newcommand{\Nn}{{\mathbb N}}

\newcommand{\Rr}{{\mathbb R}}

\newcommand{\ldeu}[2]{\cl_2\left(#1,\,#2\right)}

\newcommand{\di}[1]{{\rm d}#1}
\newcommand{\ve}[1]{\boldsymbol{#1}}

\newcommand{\ma}[1]{\boldsymbol{\rm #1}}

\newcommand{\tr}{^{\textsf T}}

\newcommand{\enu}{ , \, \dots \,,}

\newcommand{\acc}[1]{\left\{#1\right\}}

\newcommand{\Var}[1]{{\rm V}\left[ #1 \right]}

\newcommand{\Espe}[2]{{\mathbb E}_{#1}\left[#2\right]}

\newcommand{\Pro}{\mathbb{P}}
\newcommand{\Prob}[1]{{\mathbb P}\left( #1 \right)}

\newcommand{\fcar}[2] {{\mathbbm{1}}_{#1}\left(#2\right)}
\newcommand{\ie}{{\em i.e.} }
\newcommand{\eg}{{\em e.g.} }

\DeclareMathSizes{12}{10}{8}{6}

\usepackage{natbib}

\begin{document}


\title{Metamodel-based importance sampling for\\the simulation of rare events}


\author{{V. Dubourg$^{1,2}$, F. Deheeger$^{2}$, B. Sudret$^{1,2}$} \\[10pt]
\aff{$^1$ Clermont Université, IFMA, EA 3867, Laboratoire de Mécanique et Ingénieries, BP 10448} \\
\aff{\hspace*{2mm} F-63000 Clermont-Ferrand} \\
\aff{$^2$ Phimeca Engineering, Centre d'Affaires du Zénith, 34 rue de Sarliève, F-63800 Cournon d'Auvergne}}


\date{}


\abstract{In the field of structural reliability, the Monte-Carlo estimator is considered as the reference probability estimator. However, it is still untractable for real engineering cases since it requires a high number of runs of the model. In order to reduce the number of computer experiments, many other approaches known as reliability methods have been proposed. A certain approach consists in replacing the original experiment by a surrogate which is much faster to evaluate. Nevertheless, it is often difficult (or even impossible) to quantify the error made by this substitution. In this paper an alternative approach is developed. It takes advantage of the kriging meta-modeling and importance sampling techniques. The proposed alternative estimator is finally applied to a finite element based structural reliability analysis.}


\maketitle


\section{Introduction}

Reliability analysis consists in the assessment of the level of safety of a system. Given a probabilistic model (a random vector $\ve{X}$ with \emph{probability density function} (PDF) $f$) and a performance model (a function $g$), it makes use of mathematical techniques in order to estimate the system's level of safety in the form of a failure probability. A basic approach, which makes reference, is the Monte-Carlo simulation technique that resorts to numerical simulation of the performance model through the probabilistic model. Failure is usually defined as the event $F=\acc{g\left(\ve{X}\right) \leq 0}$, so that the failure probability is defined as follows:%
\begin{equation} \label{eq:pf_def}
 p_f = \Prob{F} = \Prob{\acc{g\left(\ve{X}\right) \leq 0}} = \int_{g\left(\ve{x}\right) \leq 0} f\left(\ve{x}\right)\,\di{\ve{x}}
\end{equation}%
Introducing the failure indicator function $\mathbbm{1}_{g \leq 0}$ being equal to one if $g\left(\ve{x}\right) \leq 0$ and zero otherwise, the failure probability turns out to be the mathematical expectation of this indicator function with respect to the joint probability density function $f$ of the random vector $\ve{X}$. The Monte-Carlo estimator is then derived from this convenient definition. It reads:%
\begin{equation}
 \widehat{p}_{f\,{\rm MC}} = \widehat{\mathbb{E}}_f\left[\mathbbm{1}_{g \leq 0}\left(\ve{X}\right)\right] = \frac{1}{N}\,\sum_{k=1}^N \mathbbm{1}_{g \leq 0}\left(\ve{x}^{\left(k\right)}\right)
\end{equation}%
where $\acc{x^{\left(1\right)}\enu x^{\left(N\right)}}$, $N\in\Nn^*$ is a set of samples of the random vector $\ve{X}$. According to the central limit theorem, this estimator is asymptotically unbiased and normally distributed with variance:%
\begin{equation}
 \sigma_{\widehat{p}_{f\,{\rm MC}}}^2 = \Var{\widehat{p}_{f\,{\rm MC}}} = \frac{p_f\,\left(1-p_f\right)}{N}
\end{equation}%
In practice, this variance is compared to the unbiased estimate of the failure probability in order to decide whether it is accurate enough or not. The \emph{coefficient of variation} is defined as $\delta_{\widehat{p}_{f\,{\rm MC}}} = \sigma_{\widehat{p}_{f\,{\rm MC}}}/\widehat{p}_{f\,{\rm MC}}$. Given $N$, this coefficient dramatically increases as soon as the failure event is too rare ($p_f \rightarrow 0$) and proves that the Monte-Carlo estimation technique intractable for real world engineering problems for which the performance function involves the output of an expensive-to-evaluate black box function -- \eg a finite element code. Note that this remark is also true for too frequent events ($p_f \rightarrow 1$) as the coefficient of variation of $1-\widehat{p}_{f\,{\rm MC}}$ exhibits the same property.\par

In order to reduce the number of simulation runs, a large set of other approaches known as reliability methods have been proposed. They might be classified as follows.\par

A first approach consists in replacing the original experiment by a \emph{surrogate} which is much faster to evaluate. Among such approaches, there are the well-known first and second order reliability methods \citep[\eg][]{Ditlevsen1996,Lemaire2009}, quadratic response surfaces \citep{Bucher1990} and the more recent meta-models such as \emph{support vector machines} \mbox{\citep{Hurtado2004b,Deheeger2007}}, \emph{neural networks} \citep{Papadrakakis2002} and \emph{kriging} \citep{Kaymaz2005,Bichon2008}. Nevertheless, it is often difficult or even impossible to quantify the error made by such a substitution.\par

The other approaches are the so-called \emph{variance reduction techniques}. In essence, these techniques aims at favoring the Monte-Carlo simulation of the failure event $F$ in order to reduce the estimation variance. These approaches are more robust because they do not rely on any assumption regarding the functional relationship $g$, though they are still too computationally demanding to be implemented for industrial cases. For an extended review of these techniques, the reader is referred to the book by \citet{Rubinstein2008}.\par

In this paper an hybrid approach is developed. It is based on both \emph{margin meta-models} (defined hereafter) and the importance sampling technique. It is then applied to an academic structural reliability problem involving a linear finite-element model and a two-dimensional random field.\par


\section{Adaptive probabilistic classification using \emph{margin meta-models}}

A meta-model means to a model what the model itself means to the real-world. Loosely speaking, it is \emph{the model of the model}. As opposed to the model, its construction does not rely on any physical assumption about the phenomenon of interest but rather on statistical considerations about the coherence of some scattered observations that result from a set of experiments. This set is usually referred to as a \emph{design of experiments} (DOE): $\cx = \{\ve{x}_1\enu\ve{x}_m\}$. It should be carefully selected in order to retrieve the largest amount of statistical information about the underlying functional relationship over the input space $\cd_{\ve{x}}$. Here, we attempt to build a meta-model for the failure indicator function $\mathbbm{1}_{g \leq 0}$. In the \emph{statistical learning theory} \citep{Vapnik1995} this is referred to as a \emph{classification} problem.\par

Hereafter, we define a \emph{margin meta-model} as a meta-model that is able to give a \emph{probabilistic prediction} of the response quantity of interest whose spread (\ie variance) depends on the lack of information brought by the DOE. It is thus reducible by bringing more observations into the DOE. In other words, this is an epistemic (reducible) source of uncertainty. To the authors' knowledge, there exist only two families of such margin meta-models: the probabilistic support vector machines $\mathbb{P}$-SVM by \citet{Platt1999} and Gaussian-Process- (or kriging-) based classification \citep{Santner2003}. The present paper makes use of the kriging meta-model, but the overall concept could easily be extended to $\mathbb{P}$-SVM. The theoretical aspects of the kriging prediction are briefly introduced in the following subsection before it is applied to the classification problem of interest.\par

\subsection{Gaussian-process based prediction}

The Gaussian-Process based prediction (also known as \emph{kriging}) theory is detailed in the book by \citet{Santner2003}. In essence, kriging assumes that the performance function $g$ is a sample path of an underlying Gaussian stochastic process $G$ that would read as follows:%
\begin{equation}
 G\left(\ve{x}\right) = \ve{f}\left(\ve{x}\right)\tr\,\ve{\beta} + Z\left(\ve{x}\right)
\end{equation}%
where:%
\begin{itemize}
 \item $\ve{f}\left(\ve{x}\right)\tr\,\ve{\beta}$ denotes the mean of the GP which corresponds to a classical linear regression model on a given functional basis $\acc{f_i,\;i = 1\enu p} \in \ldeu{\cd_{\ve{x}}}{\Rr}$~;
 \item $Z\left(\ve{x}\right)$ denotes the stochastic part of the GP which is modelled as a zero mean, constant variance $\sigma_G^2$, stationary Gaussian process with a given symmetric positive definite autocorrelation model. It is fully defined by its autocovariance function which reads ($\left(\ve{x},\,\ve{x}'\right) \in \cd_{\ve{x}}\times\cd_{\ve{x}}$):%
 \begin{equation}
  C_{GG}\left(\ve{x},\,\ve{x}'\right) = \sigma_G^2\,R\left(\left|\ve{x}-\ve{x}'\right|,\ve{\ell}\right)
 \end{equation}%
 where $\ve{\ell}$ is a vector of parameters defining $R$.
\end{itemize}%
The most widely used class of autocorrelation functions is the anisotropic squared exponential model:%
\begin{equation} \label{eq002}
R\left(\left|\ve{x}-\ve{x}'\right|,\ve{\ell}\right) = \exp\left(\sum\limits_{k=1}^n - \left|\frac{x_k-x_k'}{\ell_k}\right|^2\right)
\end{equation}%

The best linear unbiased estimation of $G$ at point $\ve{x}$ is shown \citep[Chap. 8]{Santner2003,Severini2005} to be the following Gaussian random variate:
\begin{equation}
  \begin{split}
    \widehat{G}\left(\ve{x}\right) &= G(\ve{x}) \left| \{g(\ve{x}_1)\enu g(\ve{x}_m)\} \right. \\[-10pt]
				   &\sim \cn\left(\mu_{\widehat{G}}\left(\ve{x}\right),\,\sigma_{\widehat{G}}\left(\ve{x}\right)\right)
  \end{split}
\end{equation}
with moments:%
\begin{align}
  \mu_{\widehat{G}}\left(\ve{x}\right) &= \ve{f}\left(\ve{x}\right)\tr\,\ve{\widehat{\beta}} + \ve{r}\left(\ve{x}\right)\tr\ma{R}^{-1}\left(\ve{Y} - \ma{F}\,\ve{\widehat{\beta}}\right)\\[-10pt]
  \begin{split}
    \sigma_{\widehat{G}}^2\left(\ve{x}\right) &= \\[-10pt] &\sigma_{G}^2 \left(1 -
      \left[\begin{array}{c}
	\ve{f}\left(\ve{x}\right)\\
	\ve{r}\left(\ve{x}\right)
      \end{array}\right]\tr\,
      \left[\begin{array}{cc}
	\ma{0} & \ma{F}\tr \\
	\ma{F} & \ma{R}
      \end{array}\right]^{-1}\,
      \left[\begin{array}{c}
	\ve{f}\left(\ve{x}\right)\\
	\ve{r}\left(\ve{x}\right)
      \end{array}\right]
    \right)
  \end{split}
\end{align}%
where we have introduced $\ve{r}$, $\ma{R}$ and $\ma{F}$ such that:%
\begin{align}
 r_i(\ve{x}) &= R\left(\left|\ve{x}-\ve{x}_i\right|,\,\ve{\ell}\right),\;i=1\enu m \\[-10pt]
 R_{ij} &= R\left(\left|\ve{x}_i-\ve{x}_j\right|,\,\ve{\ell}\right),\;i=1\enu m,\;j=1\enu m \\[-10pt]
 F_{ij} &= f_i\left(\ve{x}_j\right),\;i=1\enu p,\;j=1\enu m
\end{align}%
At this stage it can easily be proven that $\mu_{\widehat{G}}\left(\ve{x}_i\right) = g\left(\ve{x}_i\right)$ and $\sigma_{\widehat{G}}\left(\ve{x}_i\right) = 0$ for $i = 1\enu m$, thus meaning the kriging surrogate is an \emph{exact interpolator}.

Given a choice for the regression and correlation models, the optimal set of parameters $\ve{\beta}^*$, $\ve{\ell}^*$ and $\sigma_{G}^{2\,*}$ can then be inferred using the \emph{maximum likelihood principle} applied to the unique sparse observation of the GP sample path grouped in the vector $\ve{y} = \left\langle g\left(\ve{x}_i\right),\;i=1\enu m\right\rangle$. This inference problem turns into an optimization problem that can be solved analytically for both $\ve{\beta}^*$ and $\sigma_{G}^{2\,*}$ assuming $\ve{\ell}^*$ is known. Thus, the problem is solved in two steps: the maximum likelihood estimation of $\ve{\ell}^*$ is first solved by a global optimization algorithm which in turns gives the optimal values of $\ve{\beta}^*$ and $\sigma_{G}^{2\,*}$.\par

\begin{figure*}
  \subfigure[\label{fig:ProbabilisticClassificationFunction_2D}Limit-states and probability contours.]{\includegraphics[width=.5\linewidth, clip=true, trim=30 -15 30 30]{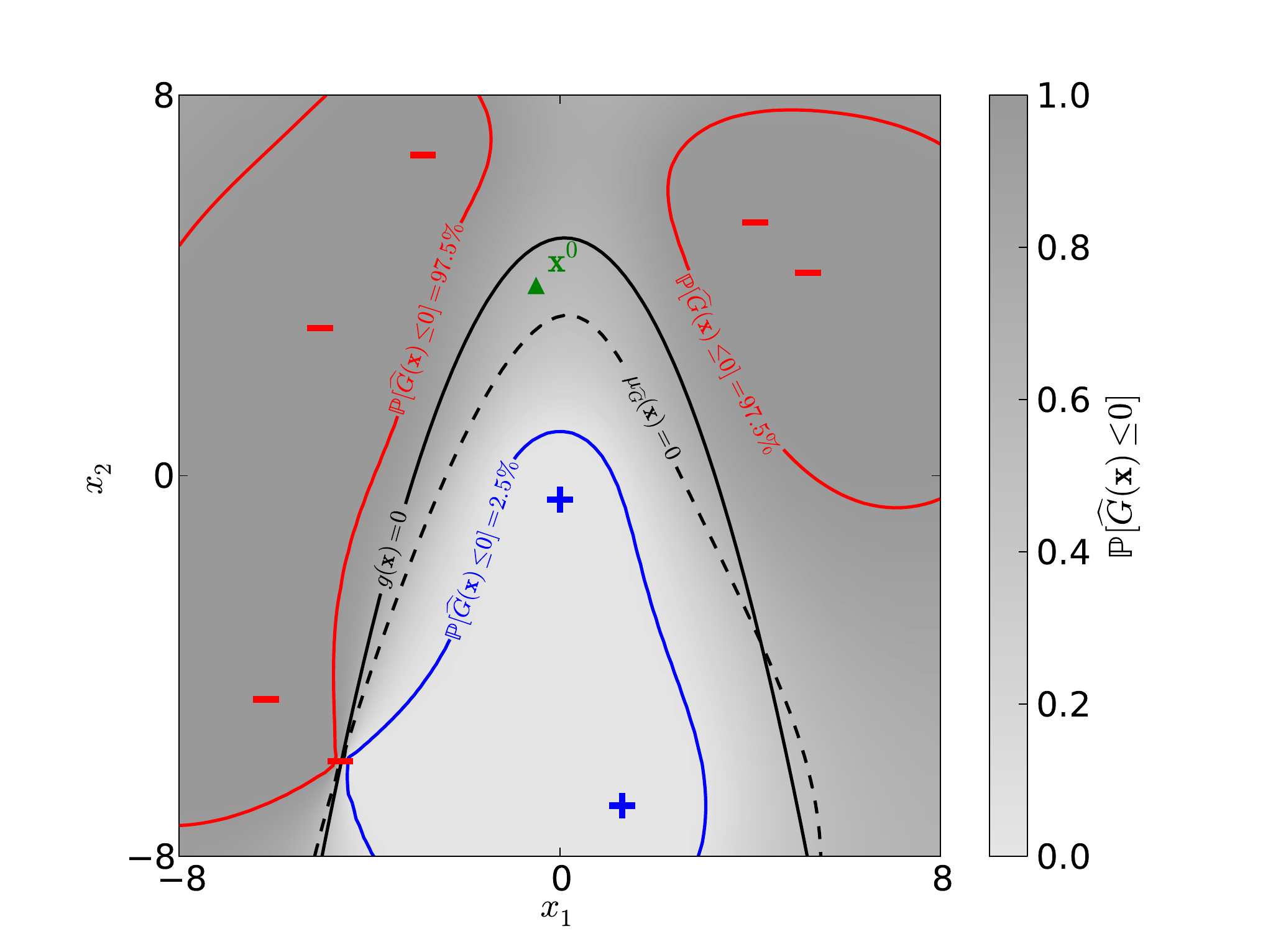}}
  \subfigure[\label{fig:ProbabilisticClassificationFunction_1D}Classification given $\ve{x}=\ve{x}^0$.]{\includegraphics[width=.5\linewidth, clip=true, trim=0 15 0 -30]{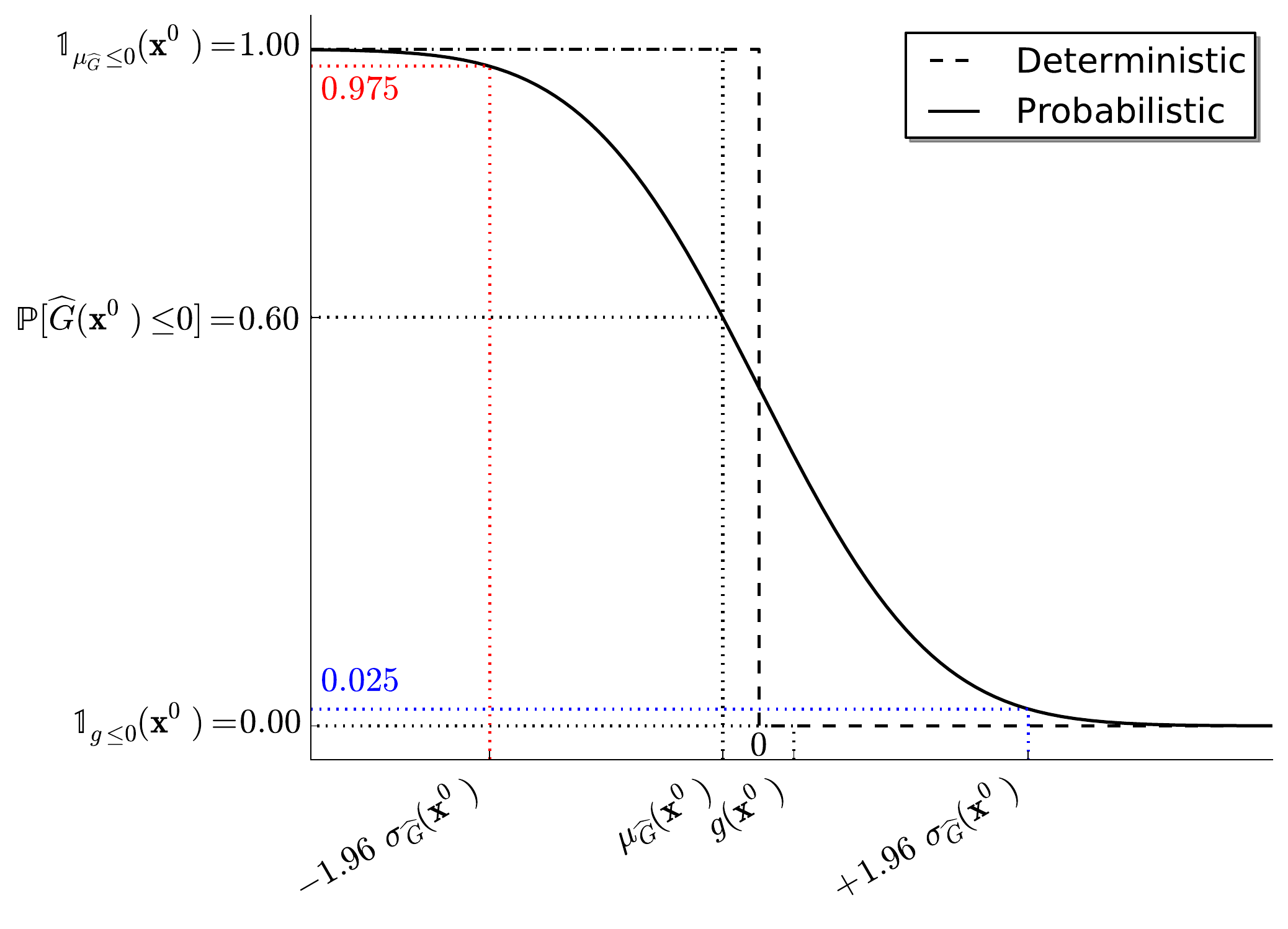}}
  \caption{Comparison of the three classification strategies on a two-dimensional example from \citet{DerKiureghian1998}.}
  \label{fig:ProbabilisticClassificationFunction}
\end{figure*}

\subsection{Probabilistic classification function} \label{sec:ProbabilisticClassificationFunction}

A surrogate-based reliability analysis simply consists in replacing the performance function $g$ by its meta-model $\mu_{\widehat{G}}$. Again, this meta-model may be a first- or second-order Taylor expansion of the limit-state surface $g=0$ at a so-called \emph{design point} \citep[FORM/SORM]{Ditlevsen1996}, a polynomial response surface \citep{Bucher1990}, a \emph{neural networks} based prediction \citep{Papadrakakis2002}, or a \emph{kriging} based prediction \citep{Bichon2008}. This surrogate may not be fully accurate and it is difficult or even impossible to quantify the error made by substitution on the final failure probability of interest.

In this paper as in the work by \citet{Picheny2009}, it is proposed to use the complete probabilistic prediction provided by the kriging meta-model instead of the sole mean prediction \citep[\eg as in][]{Bichon2008}. Indeed, since the probabilistic distribution of the prediction is fully characterized, the probability that the prediction is negative may be expressed in closed-form and reads as follows:
\begin{equation} \label{eq:ProbabilisticClassificationFunction}
  \Pro\left[\widehat{G}(\ve{x}) \leq 0\right] = \Phi\left(\frac{0-\mu_{\widehat{G}}(\ve{x})}{\sigma_{\widehat{G}}(\ve{x})}\right)
\end{equation}
Note that this latter quantity is not the sought failure probability $p_f$, this is simply the probability that the prediction $\widehat{G}$ at some deterministic vector $\ve{x}$ is negative.\par

\citet{Picheny2009} proposes then to use this probabilistic classification function as a surrogate for the real failure indicator function, and uses crude Monte-Carlo simulation to estimate the failure probability. A different use is proposed in the next section.\par

Figure \ref{fig:ProbabilisticClassificationFunction} illustrates the concepts introduced in this section on a basic structural reliability example from \citet{DerKiureghian1998}. This example involves two independent standard Gaussian random variates $X_1$ and $X_2$, and the performance function reads:
\begin{equation}
 g(x_1,\,x_2) = b - x_2 - \kappa\,(x_1 - e)^2
\end{equation}
where $b = 5$, $\kappa = 0.5$ and $e = 0.1$. In subfigure \ref{fig:ProbabilisticClassificationFunction_2D}, the limit-state function $g(\ve{x})=0$ is represented by the black dash-dot line. The red minusses ($g \leq 0$) and blue plusses ($g > 0$) represent the initial DOE from which the kriging meta-model is built. The mean prediction's limit-state $\mu_{\widehat{G}}(\ve{x})=0$ is represented by the dashed black line. It can be seen that the meta-model is not fully accurate since the green triangle $\ve{x}^0$ (among others) is misclassified. Indeed, $\ve{x}^0$ is safe according to the real performance function $g$, but it fails according to the mean prediction of the meta-model $\mu_{\widehat{G}}$. The probabilistic classification function makes a smoother decision possible: $\ve{x}^0$ fails with a 60\% probability w.r.t. the epistemic uncertainty in the random prediction $\widehat{G}(\ve{x}^0) \sim \cn(\mu_{\widehat{G}}(\ve{x}^0),\,\sigma_{\widehat{G}}(\ve{x}^0))$. Note also that the red and blue points in the DOE fails with probabilities 100\% and 0\% (safe) respectively due to the interpolating property of the kriging metamodel. Subfigure \ref{fig:ProbabilisticClassificationFunction_2D} is the one-dimensional illustration of the three classification strategies for the vector $\ve{x}^0$. The deterministic decision function is an heaviside function centered in zero, and the probabilistic classification is a smoother Gaussian cumulative density function.

\subsection{Refinement of the probabilistic classification function}\label{sec:DOE}

In this subsection, a strategy is proposed in order to refine the probabilistic classification function so that it tends towards the real indicator function $\mathbbm{1}_{g \leq 0}$.\par

First, let the \emph{margin of uncertainty} $\mathbb{M}$ be defined as follows:
\begin{equation}
  \mathbb{M} = \acc{\ve{x}\,:\,-k\,\sigma_{\widehat{G}}\left(\ve{x}\right) \leq \widehat{G}\left(\ve{x}\right) \leq +k\,\sigma_{\widehat{G}}\left(\ve{x}\right)}
\end{equation}
where $k$ might be chosen as $k=\Phi^{-1}(97.5\%)=1.96$ meaning a 95\% confidence interval onto the prediction of the limit-state surface is chosen. Such a 95\% confidence margin is illustrated in Subfigure \ref{fig:ProbabilisticClassificationFunction_2D} as the area bounded below by the blue line (2.5\% confidence level) and above by the red line (97.5\% confidence level). The points that are located in this margin have an uncertain sign, the others being either failed or safe with a confidence level greater than 97.5\%.\par

We also define the probability that a point $\ve{x} \in \cd_{\ve{x}}$ belongs to this margin of uncertainty. Due to the Gaussian nature of the prediction, this probability may also be expressed in closed-form and reads as follows:
\begin{equation}
  \begin{split}
    \Pro\left[\ve{x} \in \mathbb{M}\right]
      &= \Phi\left(\frac{k\,\sigma_{\widehat{G}}\left(\ve{x}\right) - \mu_{\widehat{G}}\left(\ve{x}\right)}{\sigma_{\widehat{G}}\left(\ve{x}\right)}\right)\\[-10pt]
      &- \Phi\left(\frac{-k\,\sigma_{\widehat{G}}\left(\ve{x}\right) - \mu_{\widehat{G}}\left(\ve{x}\right)}{\sigma_{\widehat{G}}\left(\ve{x}\right)}\right)
  \end{split}
\end{equation}\par

Then, finding the point that maximizes this quantity on the support of the PDF of $\ve{X}$ will finally bring the best improvement point in the DOE. Starting with this statement, many authors in the kriging literature decide to use global optimization algorithms in order to find \emph{the best} improvement point. For instance, \citet{Bichon2008} use a different criterion named the \emph{expected feasibility function}, and \citet{Lee2008} use the \emph{constraint boundary sampling} criterion. Note also that an equivalent concept is used by \citet{Hurtado2004b,Deheeger2007,Deheeger2008,Bourinet2010} for SVM.\par

In this paper, as in \citet{Dubourg2011}, a slightly different strategy is proposed in order to add several points in the DOE. The proposed criterion $\Pro\left[\ve{x} \in \mathbb{M}\right]$ is multiplied by a weighting density function $w$ so that
\begin{equation}
  \cc\left(\ve{x}\right) = \Prob{\ve{x}\in\mathbb{M}} \, w\left(\ve{x}\right)
\end{equation}
can itself be regarded as a PDF up to an unknown but finite normalizing constant. The weighting density $w$ can either be chosen as the original PDF of $\ve{X}$, or, as it is proposed here, the uniform PDF on a sufficiently large confidence region of the original PDF. Such a confidence region might be difficult to define for any given PDF, but as it is usually done in structural reliability \citep{Ditlevsen1996}, the original random vector $\ve{X}$ can be transformed into a probabilistically equivalent standard Gaussian random vector $\ve{U}$ for which the confidence region is simply an hypersphere with radius $\beta_0$. The reader is referred to \citet{Lebrun2009a} for a recent discussion on such mappings $\ve{U} = T(\ve{X})$. In that given space, $\beta_0$ can be easily selected as \eg $\beta_0=8$ which corresponds to the maximal \emph{generalized reliability index} \citep{Ditlevsen1996} that can be justified numerically, and the sought uniform PDF is simply defined in terms of the following indicator function:%
\begin{equation}
  w\left(\ve{u}\right) \propto \fcar{\sqrt{\ve{u}\tr\,\ve{u}} \leq \beta_0}{\ve{u}}
\end{equation}\par

Markov-chain Monte-Carlo simulation techniques \citep[\eg the \emph{slice sampling} technique proposed by][]{Neal2003} might be used in order to generate $N$ (say $N=10^4$) samples from the pseudo-PDF $\cc$. These samples are expected to be highly concentrated around the maxima of the criterion $\Prob{\ve{x}\in\mathbb{M}}$, and thus in the vicinity of the predicted limit-state $\mu_{\widehat{G}}(\ve{x})=0$ where the sign of $\widehat{G}$ is the most uncertain. This large candidate population can then be reduced to a smaller one that condensate its statistical properties by means of a $K$-means clustering algorithm \citep{MacQueen1967}. The $K$ ($K$ being given) cluster centers uniformly span the margin $\mathbb{M}$ and may be added to the DOE in order to enrich the prediction of the performance function in the vicinity of the limit-state and thus reduce the margin of uncertainty.\par


\section{Meta-model-based importance sampling}\label{sec:metaIS}

\citet{Picheny2009} proposes to use the probabilistic classification function as a surrogate for the real indicator function, so that the failure probability is rewritten from its definition in \eqref{eq:pf_def} as follows:
\begin{equation}\label{eq:AugmentedFailureProbability}
 p_{f\,\varepsilon} = \int \Pro\left[\widehat{G}(\ve{x}) \leq 0\right]\,f(\ve{x})\,\di{\ve{x}} \equiv \mathbb{E}_f\left[\Pro\left[\widehat{G}(\ve{X}) \leq 0\right]\right]
\end{equation}\par

It is argued here that this latter quantity does not equal the failure probability of interest because it sums the aleatory uncertainty in the random vector $\ve{X}$ and the epistemic uncertainty in the prediction $\widehat{G}$. This is the reason why $p_{f\,\varepsilon}$ will hereafter be referred to as the \emph{augmented failure probability}. As a matter of fact, even if the epistemic uncertainty in the prediction can be reduced (\eg by enriching the DOE as proposed in section \ref{sec:DOE}), it is impossible to quantify the contribution of each source of uncertainty \emph{a posteriori}.\par

This remark motivates the approach introduced in this section where the probabilistic classification function is used in conjunction with the importance sampling technique in order to build a new estimator of the failure probability.\par

\subsection{Importance sampling}

According to \citet{Rubinstein2008}, importance sampling (IS) is the most efficient variance reduction technique. This technique consists in computing the mathematical expectation of the failure indicator function according to a biased PDF which favors the failure event of interest. This PDF is called the \emph{instrumental density}.\par

Given an instrumental density $h$, such that $h$ dominates $\mathbbm{1}_{g\leq0}\,f$, the definition of the failure probability of \eqref{eq:pf_def} may be rewritten as follows:
\begin{equation}\label{eq:pfIS_def}
  p_f = \int_{g(\ve{x}) \leq 0} \frac{f(\ve{x})}{h(\ve{x})}\,h(\ve{x})\,\di{\ve{x}} \equiv \Espe{h}{\mathbbm{1}_{g \leq 0}(\ve{X}) \frac{f(\ve{X})}{h(\ve{X})}}
\end{equation}
where the subscript $h$ on the expectation operator is added to recall that $\ve{X}$ is therefore distributed according to $h$. Note that the domination requirement of $h$ over $\mathbbm{1}_{g\leq0}\,f$ simply means that:
\begin{equation}
  \forall \ve{x}\in\cd_{\ve{x}},\quad h(\ve{x}) = 0 \Rightarrow \mathbbm{1}_{g\leq0}(\ve{x})\,f(\ve{x})=0
\end{equation}
so that the so-called \emph{likelihood ratio} $\ell(\ve{x}) = f(\ve{x})/h(\ve{x})$ is finite for any given $\ve{x}\in\cd_{\ve{x}}$.\par

The latter definition of the failure probability easily leads to the establishment of the importance sampling estimator which reads as follows:
\begin{equation}
  \widehat{p}_{f\,{\rm IS}} = \widehat{\mathbb{E}}_f\left[\mathbbm{1}_{g \leq 0}\left(\ve{X}\right)\right] = \frac{1}{N}\,\sum_{k=1}^N \mathbbm{1}_{g \leq 0}\left(\ve{x}^{\left(k\right)}\right)\,\frac{f(\ve{x}^{\left(k\right)})}{h(\ve{x}^{\left(k\right)})}
\end{equation}
where $\acc{x^{\left(1\right)}\enu x^{\left(N\right)}}$, $N\in\Nn^*$ is a set of samples from $h$. According to the central limit theorem, this estimation is unbiased and its quality may be measured by means of its variance of estimation which reads:
\begin{equation}
  \begin{split}
    {\rm Var}\left[\widehat{p}_{f\,{\rm IS}}\right] &=\\[-10pt] &\hspace*{-12mm}\frac{1}{N-1}\,\left(\frac{1}{N}\,\sum\limits_{k=1}^N \mathbbm{1}_{g \leq 0}(\ve{x}^{(k)})\,\left(\frac{f(\ve{x}^{(k)})}{h(\ve{x}^{(k)})}\right)^2 -  \widehat{p}_{f\,{\rm IS}}^2\right)
  \end{split}
\end{equation}\par

\citet{Rubinstein2008} show that this variance is zero (optimality of the IS estimator) when the instrumental PDF is chosen as:
\begin{equation} \label{eq:optISdens}
 h^*(\ve{x}) = \frac{\mathbbm{1}_{g \leq 0}(\ve{x})\,f(\ve{x})}{\int \mathbbm{1}_{g \leq 0}(\ve{x})\,f(\ve{x})\,\di{\ve{x}}} = \frac{\mathbbm{1}_{g \leq 0}(\ve{x})\,f(\ve{x})}{p_f}
\end{equation}
However this instrumental PDF is not implementable in practice because it involves the sought failure probability in its denominator. There exists infinitely many PDF $h$ that allows to significantly reduce the variance of estimation though.\par

\subsection{A meta-model-based approximation of the optimal instrumental PDF}

Different strategies have been proposed in order to build quasi-optimal instrumental PDF suited for specific estimation problems. For instance, \citet{Melchers1989} uses a standard normal PDF centered onto the \textit{most probable failure point} (MPFP) in the space of the independent standard Gaussian random variables $\ve{U}=T(\ve{X})$ in order to estimate a failure probability. Although this approach may lose accuracy as soon as the MPFP is not unique. \citet{Cannamela2008} use a kriging prediction of the performance function $g$ in order to build an instrumental PDF suited for the estimation of extreme quantiles of the random variate $G=g(\ve{X})$.\par

\begin{figure*}
  \subfigure[\label{fig:OptimalInstrumentalPDF}The optimal instrumental PDF.]{\includegraphics[width=.5\linewidth, clip=true, trim=30 25 30 30]{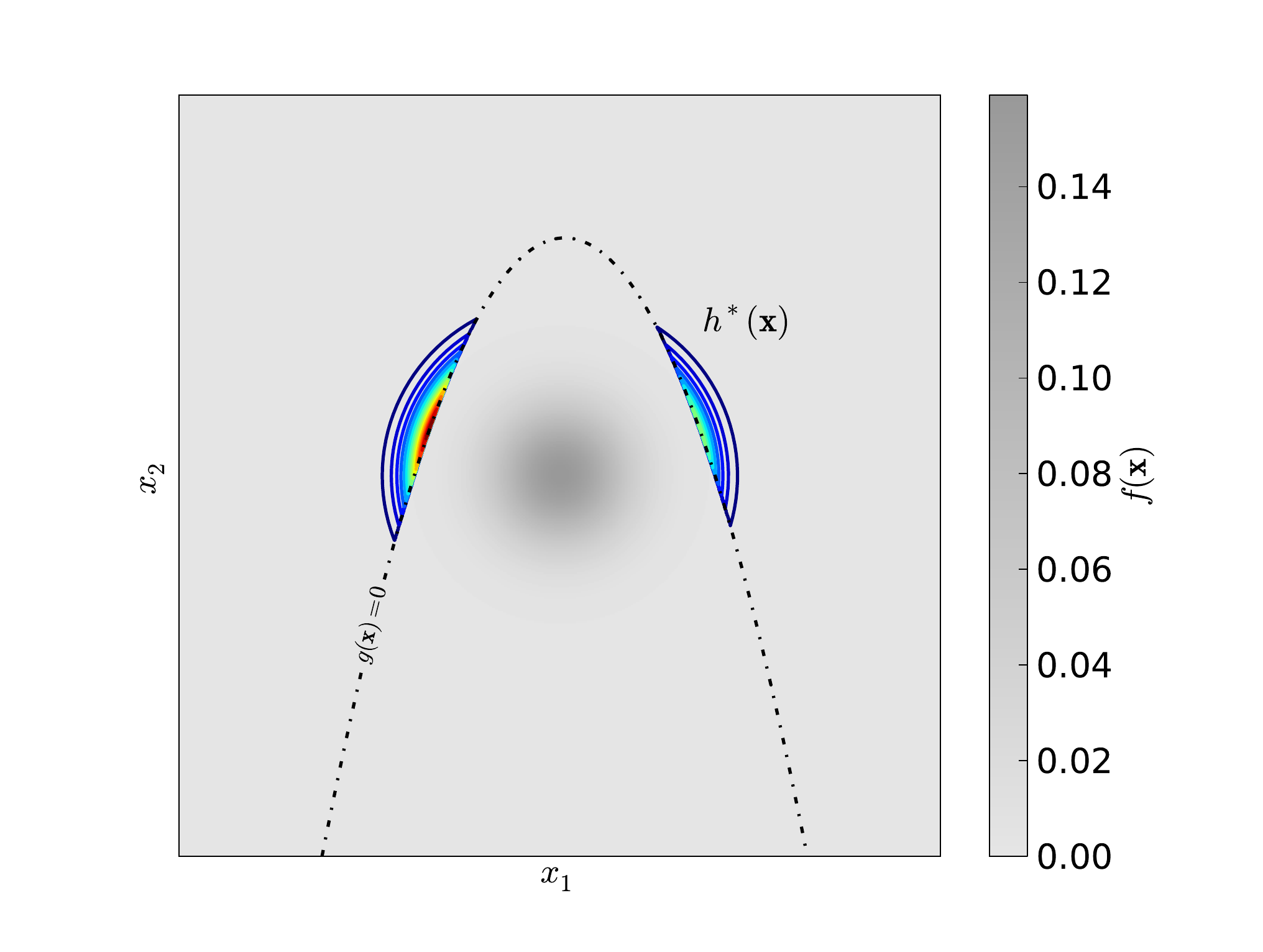}}
  \subfigure[\label{fig:QuasiOptimalInstrumentalPDF}A quasi-optimal PDF.]{\includegraphics[width=.5\linewidth, clip=true, trim=30 25 30 30]{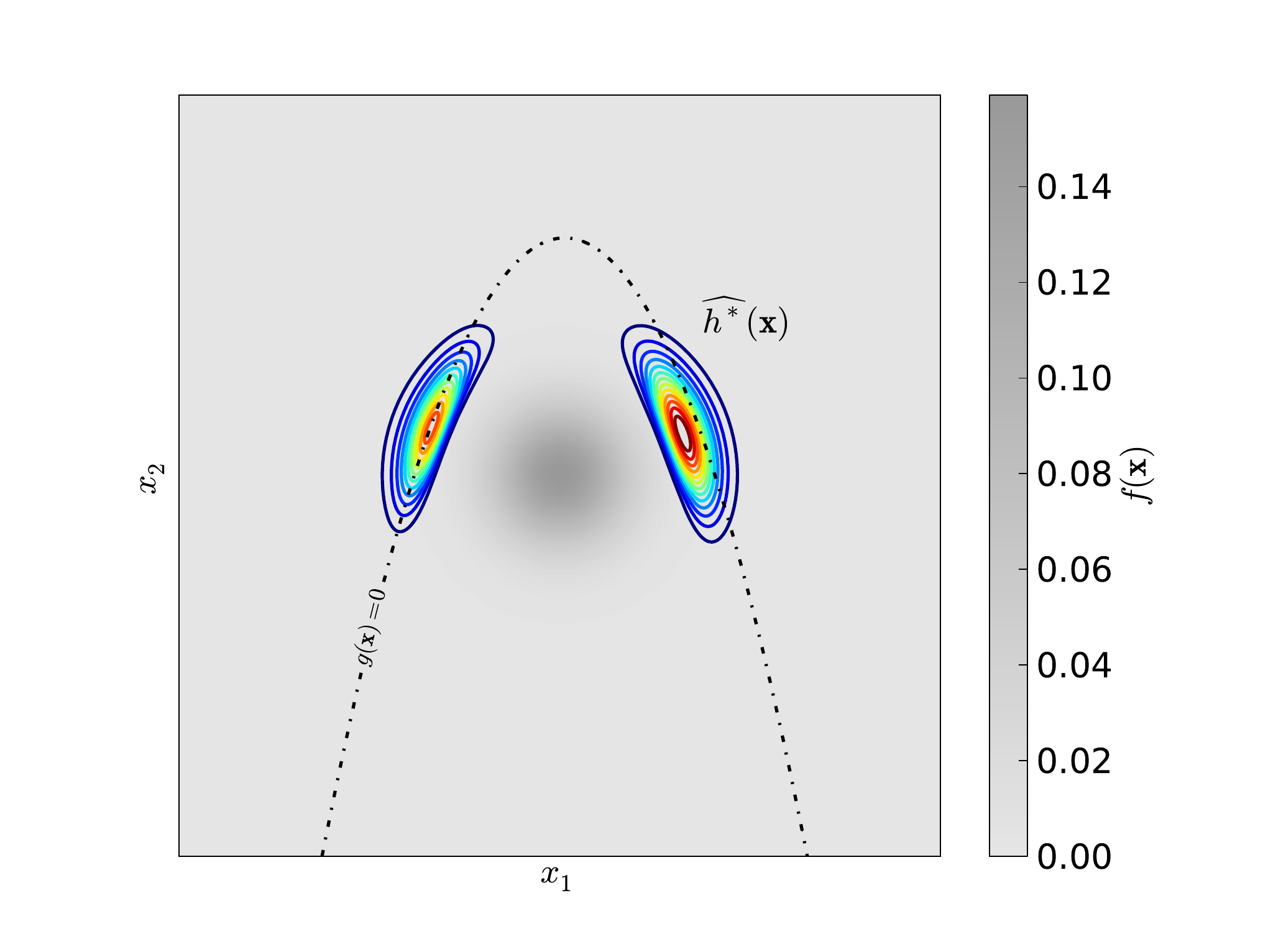}}
  \caption{Comparison of the instrumental PDF on the two-dimensional example from \citet{DerKiureghian1998}.}
  \label{fig:InstrumentalPDFComparison}
\end{figure*}

Here, it is proposed to use the probabilistic classification function in \eqref{eq:ProbabilisticClassificationFunction} as a surrogate for the real indicator function in the optimal instrumental PDF in \eqref{eq:optISdens}. The proposed quasi-optimal PDF thus reads as follows:
\begin{equation} \label{eq:quasioptISdens}
  \widehat{h^{*}}(\ve{x}) = \frac{\Pro\left[\widehat{G}(\ve{x}) \leq 0\right]\,f(\ve{x})}{\int \Pro\left[\widehat{G}(\ve{x}) \leq 0\right]\,f(\ve{x})\,\di{\ve{x}}} \equiv \frac{\mathbbm{1}_{\widehat{G} \leq 0}(\ve{x})\,f(\ve{x})}{p_{f\,\varepsilon}}
\end{equation}
where $p_{f\,\varepsilon}$ is the augmented failure probability which has been already defined in \eqref{eq:AugmentedFailureProbability}. This quasi-optimal instrumental PDF is compared to its impractical optimal counterpart in Figure \ref{fig:InstrumentalPDFComparison} using the example of Section \ref{sec:ProbabilisticClassificationFunction}.

\subsection{The meta-model-based importance sampling estimator}\label{sec:pf_metaIS}

Choosing the proposed quasi-optimal instrumental PDF in \eqref{eq:quasioptISdens} in the importance sampling definition of the failure probability in \eqref{eq:pfIS_def} leads to the following new definition:
\begin{align}
  p_f &= \int \mathbbm{1}_{g \leq 0}(\ve{x}) \frac{f(\ve{x})}{\widehat{h^*}(\ve{x})}\,\widehat{h^*}(\ve{x})\,\di{\ve{x}} \\
      &= p_{f\,\varepsilon} \, \int \frac{\mathbbm{1}_{g \leq 0}(\ve{x})}{\Pro\left[\widehat{G}(\ve{x}) \leq 0\right]}\,\widehat{h^*}(\ve{x})\,\di{\ve{x}} \\
      &\equiv p_{f\,\varepsilon} \, \alpha_{\rm corr} \\
  \intertext{where we have introduced:}
  \alpha_{\rm corr} &\equiv \Espe{\widehat{h^*}}{\frac{\mathbbm{1}_{g \leq 0}(\ve{X})}{\Pro\left[\widehat{G}(\ve{X}) \leq 0\right]}} \label{eq:pfcorr}
\end{align}
This means that the failure probability is now defined as the product between the augmented failure probability $p_{f\,\varepsilon}$ and a correction factor $\alpha_{\rm corr}$. This correction factor is defined as the expected ratio between the real indicator function $\mathbbm{1}_{g \leq 0}$ and the probabilistic classification function $\Pro[\widehat{G}(\bullet) \leq 0]$. Thus, if the kriging prediction is fully accurate, the correction factor is equal to unity and the failure probability is identical to the augmented failure probability (optimality of the proposed estimator). On the other hand, in the more general case where the kriging prediction is not fully accurate, the correction factor modifies the augmented failure probability accounting for the epistemic uncertainty in the prediction.\par

The two terms of the latter definition of the failure probability may now be estimated using Monte-Carlo simulation:
\begin{align}
  \widehat{p}_{f\,\varepsilon} &= \frac{1}{N_{\varepsilon}} \sum\limits_{k=1}^{N_{\varepsilon}} \Pro\left[\widehat{G}(\ve{x}^{(k)}) \leq 0\right] \\[-10pt]
  \widehat{\alpha}_{\rm corr} &= \frac{1}{N_{\rm corr}} \sum\limits_{k=1}^{N_{\rm corr}} \frac{\mathbbm{1}_{g \leq 0}(\ve{x}^{(k)})}{\Pro\left[\widehat{G}(\ve{x}^{(k)}) \leq 0\right]}
\end{align}%
where the first $N_{\varepsilon}$-sample is generated from the original PDF $f$, and the second $N_{\rm corr}$-sample is generated from the quasi-optimal instrumental PDF $\widehat{h^*}$. According to the central limit theorem, these two estimates are unbiased and normally distributed. Their respective variance of estimation denoted by $\sigma_{\varepsilon}^2$ and $\sigma_{\rm corr}^2$ are not given here but they might be easily derived.\par

To generate samples from $\widehat{h^*}$, it is proposed to use a Markov chain Monte-Carlo simulation technique which is applicable to a broad class of \emph{improper PDF} for which the normalizing constant is not known and thus for the instrumental PDF of interest $\widehat{h^*}(\ve{x}) \propto \Pro[\widehat{G}(\ve{x}) \leq 0]\,f(\ve{x})$. The work presented here makes use of the slice sampling technique \citep{Neal2003}.\par

Finally, the final estimator of the failure probability simply reads as follows:
\begin{equation} \label{eq:pf_metaIS_est}
  \widehat{p}_{f\,{\rm metaIS}} = \widehat{p}_{f\,\varepsilon} \, \widehat{\alpha}_{\rm corr}
\end{equation}
The calculation of the coefficient of variation of the final estimator $\widehat{p}_{f\,{\rm metaIS}}$ will be detailed in a forthcoming paper.


\section{Reliability analysis of an 8-hole plate}

\begin{figure*}
  \subfigure[\label{fig:8holeplate_mesh}Mesh, boundary conditions and loads.]{
    \begin{tikzpicture}[scale=.85]
      \node(0,0) {\includegraphics[width=6.8cm]{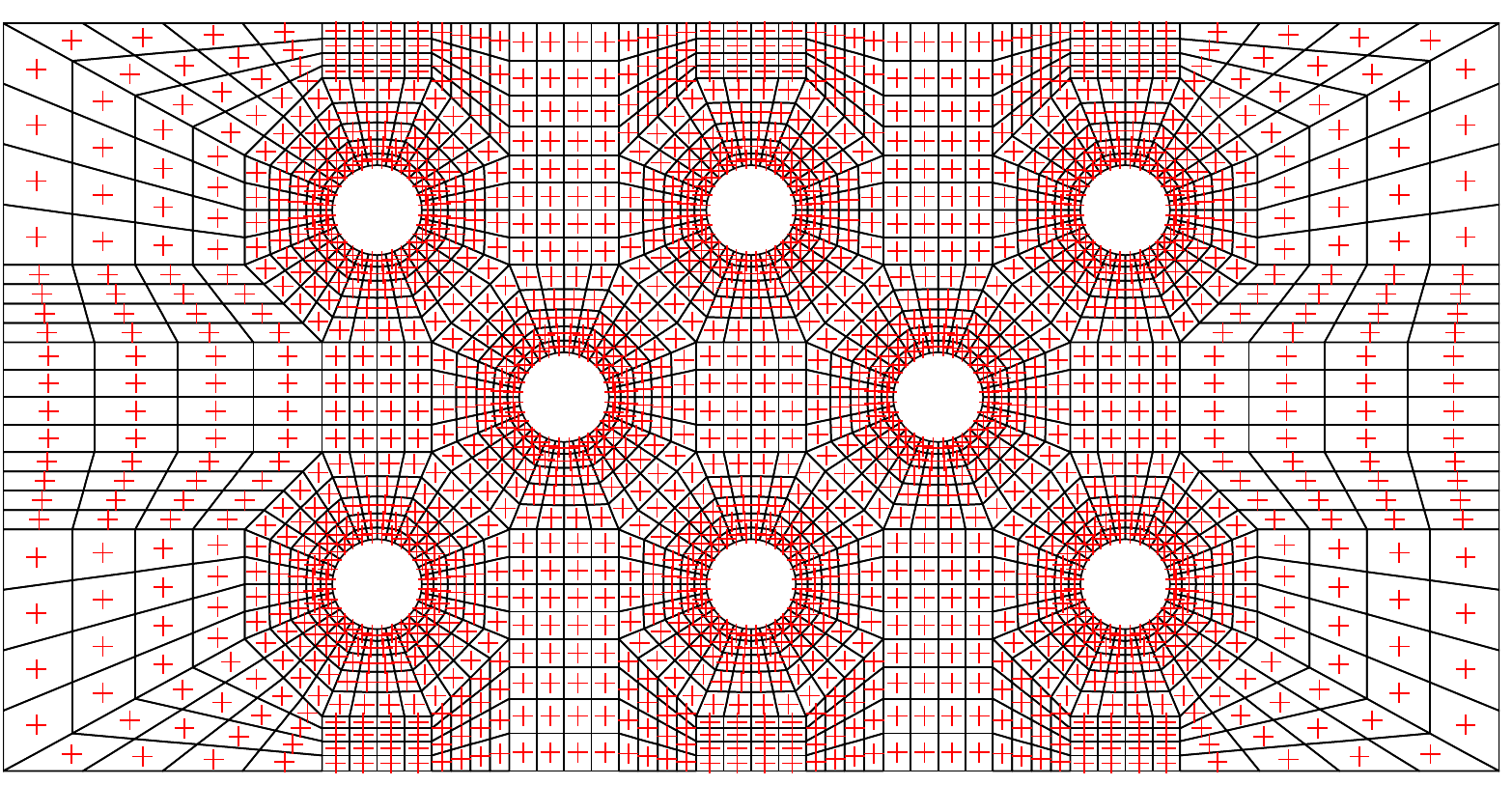}};
      \draw [very thick] (-4,-2) -- (-4,2);
      \node (d) at (-4.5,0) [above,rotate = 90] {Clamped};
      \fill [pattern = north east lines] (-4,-2) -- (-4,2) -- (-4.5,2) -- (-4.5,-2) -- cycle;
      \foreach \k in {-2,-1.5,...,2}
	{\draw [->,>=stealth] (4,\k) -- (4.5,\k);}
      \node (d) at (4.5,0) [below,rotate = 90] {$q = 100$~MPa};
    \end{tikzpicture}}
  \subfigure[\label{fig:8holeplate_mesh_randomfield_realization}One sample path of the Young modulus random field.]{
    \begin{tikzpicture}[scale=.85]
      \node(0,0) {\includegraphics[width=6.8cm]{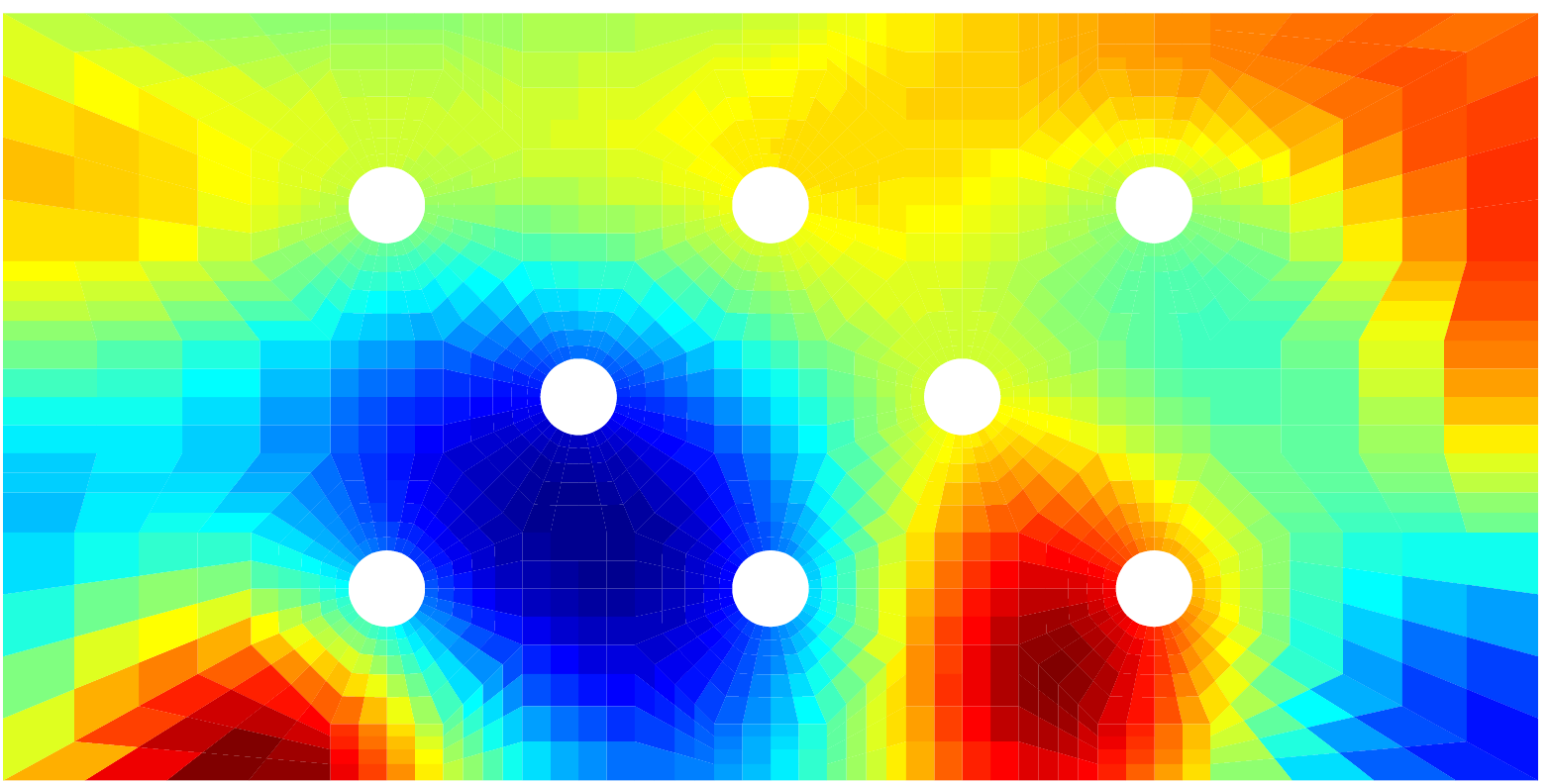}};
      \draw [very thick] (-4,-2) -- (-4,2);
      \node (d) at (-4.5,0) [above,rotate = 90] {Clamped};
      \fill [pattern = north east lines] (-4,-2) -- (-4,2) -- (-4.5,2) -- (-4.5,-2) -- cycle;
      \foreach \k in {-2,-1.5,...,2}
	{\draw [->,>=stealth] (4,\k) -- (4.5,\k);}
      \node (d) at (4.5,0) [below,rotate = 90] {$q = 100$~MPa};
    \end{tikzpicture}}
  \caption{Illustration of the 8-hole plate example from \citet{Deheeger2007}.}
  \label{fig:8holeplate}
\end{figure*}

\begin{table*}
  \begin{footnotesize}
    \renewcommand{\arraystretch}{1.2}
    \begin{center}
      \begin{tabular}{cccccc}
	\hlineB \rowcolor[gray]{.9}
				  & \textbf{DOE} & \textbf{MPFP} & \textbf{Simulations} &   \textbf{$P_f$ estimate}   & \textbf{C. o. V.} \\
	\hlineB
	\textbf{Subset (ref.)}    &      -      &      -       &   25\,000    & 1.70$\times$10$^{-5}$ &   15\%   \\
	\hline
	\textbf{multi-FORM}       &      -      &    1\,168    &       -      & 0.65$\times$10$^{-5}$ &     -    \\
	\hline
	\textbf{meta-IS}
				  &   1\,000    &     -        &       250    & 1.41$\times$10$^{-5}$ &  <~10\%   \\
	\hlineB
	\\[-10pt]
      \end{tabular}
    \end{center}
  \end{footnotesize}
  \caption{Reliability analyses results for the 8-hole plate example from \citet{Deheeger2007}.}
  \label{tab:8holeplate}
\end{table*}

This structural reliability example is inspired from \citet{Deheeger2007}. It concerns the reliability analysis of a $200\times100$~mm 8-hole plate illustrated in Figure \ref{fig:8holeplate}. The diameter of the holes $\varnothing$ is set equal to 10~mm. Its left end is clamped both horizontally and vertically while its right end is subjected to a distributed line load with magnitude $q=100$~MPa. Plain stress is assumed and the material is supposed to have a linear elastic behavior. The Poisson coefficient $\nu$ is set equal to 0.3. Due to the boundary conditions the Poisson effect is not the same on all the plate though. The Young's modulus is modeled by an homogeneous lognormal random field with a mean $\mu_E=200\,000$~MPa, a coefficient of variation $\delta_E=25\%$ and assuming an isotropic squared exponential autocorrelation function with a 20~mm correlation length $\ell$. The two-dimensional random field is represented by a \emph{translated} Karhunen-Loeve expansion discretized by means of a wavelet-Galerkin strategy proposed by \citet{Phoon2002}. The stochastic model involves 20 independent standard Gaussian random variates grouped in the vector $\ve{X}$ to simulate the random field. The mechanical model is solved with Code\_Aster \citep{CodeAster} in order to retrieve the maximal Von Mises stress in the plate $\cp$. The performance function is then defined as follows:
\begin{equation}
  g(\ve{x}) = \sigma_0 - \max\limits_{\ve{p} \in \cp}\acc{\sigma_{\rm Von~Mises}(\ve{p})}
\end{equation}
with respect to an arbitrary threshold $\sigma_0 = 450$~MPa.\par

The proposed meta-model-based importance sampling procedure is applied to this structural reliability example. First an initial kriging predictor is built for the performance function $g$ using a 100-point DOE. These 100 points are uniformally generated within the $\beta_0$ radius hypersphere. Based on this initial prediction, the DOE refinement procedure introduced in Section \ref{sec:DOE} is used. $K=100$ new points are added at each refinement iteration. The refinement procedure is stopped after 1\,000 estimations of the performance function. This may seem arbitrary but it is difficult to provide another stopping criterion for the refinement procedure -- this needs further investigation. Then, the probabilistic classification function is defined with respect to the latest (finest) kriging prediction and it is used to compute the proposed estimator of the failure probability.\par

The results are provided in Table \ref{tab:8holeplate}. They are compared to a reference solution obtained by subset simulation \citep{Au2001}, and the multi-FORM estimator from \citet{DerKiureghian1998} using FERUM v4.0 \citep{Bourinet2009} implementations of these algorithms. FERUM is a Matlab toolbox for reliability analysis published under the General Public License. The estimate of the augmented failure probability is equal to $\widehat{p}_{f\,\varepsilon} = 2.85\times10^{-5}$, and the correction factor is equal to $\widehat{\alpha}_{\rm corr} = 0.412$. It means that the kriging predictor is rather accurate in that case. The probabilistic classification function is very close to its deterministic counterpart -- and so is the instrumental importance sampling density $\widehat{h}$.\par


\section{Conclusion}

Starting from the double premise that a surrogate-based reliability analyses does not permit to quantify the substitution error, and that the existing variance reduction techniques remain time-consuming when the performance function involves the output of an expensive-to-evaluate black box function, an hybrid strategy has been proposed. First, the probabilistic classification function was introduced, this function allows a smoother classification than its deterministic counterpart accounting for the epistemic uncertainty in the kriging prediction. Using this smoother classification function within an importance sampling framework then allowed to derive a meta-model-based importance sampling estimator. This estimator converges towards the theoretically impractical optimal importance sampling estimator and may provide a significant reduction of the estimation variance as illustrated in the example.\par

In the present paper, the refinement procedure that leads to the probabilistic classification function is stopped arbitrarily. Work is in progress in order to establish the best trade-off between the size of the DOE and the number of simulations required to estimate the correction factor $\alpha_{\rm corr}$.\par


\section*{Acknowledgements}

The first author is funded by a CIFRE grant from Phimeca Engineering S.A. subsidized by the ANRT (convention number 706/2008). The financial support from the ANR through the KidPocket project is also gratefully acknowledged.

\bibliographystyle{chicaco}

\begin{thebibliography}{}

    \bibitem[\protect\citeauthoryear{Au \& Beck}{Au and Beck}{2001}]{Au2001}
    Au, S. \& J.~Beck (2001).
    \newblock Estimation of small failure probabilities in high dimensions by
    subset simulation.
    \newblock {\em Prob. Eng. Mech.\/}~{\em 16\/}(4), 263--277.

    \bibitem[\protect\citeauthoryear{Bichon, Eldred, Swiler, Mahadevan, \&
    McFarland}{Bichon et~al.}{2008}]{Bichon2008}
    Bichon, B., M.~Eldred, L.~Swiler, S.~Mahadevan, \& J.~McFarland (2008).
    \newblock {Efficient global reliability analysis for nonlinear implicit
    performance functions}.
    \newblock {\em AIAA Journal\/}~{\em 46\/}(10), 2459--2468.

    \bibitem[\protect\citeauthoryear{Bourinet, Deheeger, \& Lemaire}{Bourinet
    et~al.}{2010}]{Bourinet2010}
    Bourinet, J.-M., F.~Deheeger, \& M.~Lemaire (2010).
    \newblock Assessing small failure probabilities by combined subset simulation
    and support vector machines.
    \newblock {\em Submitted to Structural Safety\/}.

    \bibitem[\protect\citeauthoryear{Bourinet, Mattrand, \& Dubourg}{Bourinet
    et~al.}{2009}]{Bourinet2009}
    Bourinet, J.-M., C.~Mattrand, \& V.~Dubourg (2009).
    \newblock {A review of recent features and improvements added to FERUM
    software}.
    \newblock In {\em Proc. ICOSSAR'09, Int Conf. on Structural Safety And
    Reliability, Osaka, Japan}.

    \bibitem[\protect\citeauthoryear{Bucher \& Bourgund}{Bucher and
    Bourgund}{1990}]{Bucher1990}
    Bucher, C. \& U.~Bourgund (1990).
    \newblock {A fast and efficient response surface approach for structural
    reliability problems}.
    \newblock {\em Structural Safety\/}~{\em 7\/}(1), 57--66.

    \bibitem[\protect\citeauthoryear{Cannamela, Garnier, \& Iooss}{Cannamela
    et~al.}{2008}]{Cannamela2008}
    Cannamela, C., J.~Garnier, \& B.~Iooss (2008).
    \newblock {Controlled stratification for quantile estimation}.
    \newblock {\em Annals of Applied Statistics\/}~{\em 2\/}(4), 1554--1580.

    \bibitem[\protect\citeauthoryear{Deheeger}{Deheeger}{2008}]{Deheeger2008}
    Deheeger, F. (2008).
    \newblock {\em {Couplage m\'ecano-fiabiliste, $^2$SMART m\'ethodologie
    d'apprentissage stochastique en fiabilit\'e}}.
    \newblock Ph.\ D. thesis, {Universit\'e Blaise Pascal - Clermont II}.

    \bibitem[\protect\citeauthoryear{Deheeger \& Lemaire}{Deheeger and
    Lemaire}{2007}]{Deheeger2007}
    Deheeger, F. \& M.~Lemaire (2007).
    \newblock {Support vector machine for efficient subset simulations: 2SMART
    method}.
    \newblock In {\em Proc. 10th~Int. Conf. on Applications of Stat. and Prob. in
    Civil Engineering (ICASP10), Tokyo, Japan}.

    \bibitem[\protect\citeauthoryear{Der~Kiureghian \& Dakessian}{Der~Kiureghian
    and Dakessian}{1998}]{DerKiureghian1998}
    Der~Kiureghian, A. \& T.~Dakessian (1998).
    \newblock {Multiple design points in first and second-order reliability}.
    \newblock {\em Structural Safety\/}~{\em 20\/}(1), 37--49.

    \bibitem[\protect\citeauthoryear{Ditlevsen \& Madsen}{Ditlevsen and
    Madsen}{1996}]{Ditlevsen1996}
    Ditlevsen, O. \& H.~Madsen (1996).
    \newblock {\em {Structural reliability methods}\/} ({Internet (v2.3.7,
    June-Sept 2007)} ed.).
    \newblock John Wiley \& Sons Ltd, Chichester.

    \bibitem[\protect\citeauthoryear{Dubourg, Sudret, \& Bourinet}{Dubourg
    et~al.}{2011}]{Dubourg2011}
    Dubourg, V., B.~Sudret, \& J.-M. Bourinet (2011).
    \newblock Reliability-based design optimization using kriging and subset
    simulation.
    \newblock {\em Struct. Multidisc. Optim.\/}~{\em Accepted}.

    \bibitem[\protect\citeauthoryear{{eDF, R\&D Division}}{{eDF, R\&D
    Division}}{2006}]{CodeAster}
    {eDF, R\&D Division} (2006).
    \newblock {\em {Code\_Aster}~: {A}nalyse des structures et thermo-m\'ecanique
    pour des \'etudes et des recherches, V.7}.
    \newblock \textsf{http://www.code-aster.org}.

    \bibitem[\protect\citeauthoryear{Hurtado}{Hurtado}{2004}]{Hurtado2004b}
    Hurtado, J. (2004).
    \newblock {\em Structural reliability -- Statistical learning perspectives},
    Volume~17 of {\em Lecture notes in applied and computational mechanics}.
    \newblock Springer.

    \bibitem[\protect\citeauthoryear{Kaymaz}{Kaymaz}{2005}]{Kaymaz2005}
    Kaymaz, I. (2005).
    \newblock Application of kriging method to structural reliability problems.
    \newblock {\em Structural Safety\/}~{\em 27\/}(2), 133--151.

    \bibitem[\protect\citeauthoryear{Lebrun \& Dutfoy}{Lebrun and
    Dutfoy}{2009}]{Lebrun2009a}
    Lebrun, R. \& A.~Dutfoy (2009).
    \newblock An innovating analysis of the {N}ataf transformation from the copula
    viewpoint.
    \newblock {\em Prob. Eng. Mech.\/}~{\em 24\/}(3), 312--320.

    \bibitem[\protect\citeauthoryear{Lee \& Jung}{Lee and Jung}{2008}]{Lee2008}
    Lee, T. \& J.~Jung (2008).
    \newblock {A sampling technique enhancing accuracy and efficiency of
    metamodel-based RBDO: Constraint boundary sampling}.
    \newblock {\em Computers \& Structures\/}~{\em 86\/}(13-14), 1463--1476.

    \bibitem[\protect\citeauthoryear{Lemaire}{Lemaire}{2009}]{Lemaire2009}
    Lemaire, M. (2009).
    \newblock {\em Structural Reliability}.
    \newblock John Wiley \& Sons Inc.

    \bibitem[\protect\citeauthoryear{MacQueen}{MacQueen}{1967}]{MacQueen1967}
    MacQueen, J. (1967).
    \newblock Some methods for classification and analysis of multivariate
    observations.
    \newblock In J.~Le~Cam, L.M. \&~Neyman (Ed.), {\em Proc. 5$^{th}$ Berkeley
    Symp. on Math. Stat. \& Prob.}, Volume~1, Berkeley, CA, pp.\  281--297.
    University of California Press.

    \bibitem[\protect\citeauthoryear{Melchers}{Melchers}{1989}]{Melchers1989}
    Melchers, R. (1989).
    \newblock {Importance sampling in structural systems}.
    \newblock {\em {Structural Safety}\/}~{\em 6\/}(1), 3--10.

    \bibitem[\protect\citeauthoryear{Neal}{Neal}{2003}]{Neal2003}
    Neal, R. (2003).
    \newblock Slice sampling.
    \newblock {\em Annals Stat.\/}~{\em 31}, 705--767.

    \bibitem[\protect\citeauthoryear{Papadrakakis \& Lagaros}{Papadrakakis and
    Lagaros}{2002}]{Papadrakakis2002}
    Papadrakakis, M. \& N.~Lagaros (2002).
    \newblock {Reliability-based structural optimization using neural networks and
    Monte Carlo simulation}.
    \newblock {\em Comput. Methods Appl. Mech. Engrg.\/}~{\em 191\/}(32),
    3491--3507.

    \bibitem[\protect\citeauthoryear{Phoon, Huang, \& Quek}{Phoon
    et~al.}{2002}]{Phoon2002}
    Phoon, K., S.~Huang, \& S.~Quek (2002).
    \newblock Simulation of second-order processes using {Karhunen-Lo{\`{e}}ve}
    expansion.
    \newblock {\em Computers \& Structures\/}~{\em 80\/}(12), 1049--1060.

    \bibitem[\protect\citeauthoryear{Picheny}{Picheny}{2009}]{Picheny2009}
    Picheny, V. (2009).
    \newblock {\em Improving accuracy and compensating for uncertainty in surrogate
    modeling}.
    \newblock Ph.\ D. thesis, {University of Florida}.

    \bibitem[\protect\citeauthoryear{Platt}{Platt}{1999}]{Platt1999}
    Platt, J. (1999).
    \newblock Probabilistic outputs for support vector machines and comparisons to
    regularized likelihood methods.
    \newblock In {\em Advances in large margin classifiers}, pp.\  61--74. MIT
    Press.

    \bibitem[\protect\citeauthoryear{Rubinstein \& Kroese}{Rubinstein and
    Kroese}{2008}]{Rubinstein2008}
    Rubinstein, R. \& D.~Kroese (2008).
    \newblock {\em Simulation and the Monte Carlo method}.
    \newblock {Wiley Series in Probability and Statistics}. Wiley.

    \bibitem[\protect\citeauthoryear{Santner, Williams, \& Notz}{Santner
    et~al.}{2003}]{Santner2003}
    Santner, T., B.~Williams, \& W.~Notz (2003).
    \newblock {\em {The design and analysis of computer experiments}}.
    \newblock Springer series in Statistics. Springer.

    \bibitem[\protect\citeauthoryear{Severini}{Severini}{2005}]{Severini2005}
    Severini, T. (2005).
    \newblock {\em Elements of distribution theory}.
    \newblock {Cambridge series in Statistical and Probabilistic mathematics}.
    Cambridge University Press.

    \bibitem[\protect\citeauthoryear{Vapnik}{Vapnik}{1995}]{Vapnik1995}
    Vapnik, V. (1995).
    \newblock {\em {The nature of statistical learning theory}}.
    \newblock Springer.

\end{thebibliography}


\end{document}